\documentclass[usenatbib]{mn2e}
\RequirePackage{graphicx}
\begin{document}
\title{The prestellar and protostellar population of R Coronae Australis}
\author[Nutter, Ward-Thompson \& Andr\'{e}]
{David J. Nutter$^1$\thanks{E-mail: David.Nutter@astro.cf.ac.uk}, 
D. Ward-Thompson$^1$,
P. Andr\'{e}$^2$ \\
$^1$Department of Physics and Astronomy, University of Wales Cardiff,
PO Box 913, Cardiff, CF24 3YB, UK\\ 
$^2$CEA, DSM, DAPNIA, Service d'Astrophyique, C.E. Saclay, F-91191 Gif-sur-Yvette Cedex, France}

\maketitle

\begin{abstract}
We present 450 and 850~$\mu$m maps of R Coronae Australis. We compare the maps to previous surveys of the region, and shed new light on the previously unknown nature of the protostellar sources at the centre of the cloud. We clarify the nature of two millimetre sources previously discovered in lower resolution data. We identify one new Class 0 protostar that we label SMM~1B, and we measure the envelope masses of a number of more evolved protostars. We identify two new prestellar cores that we call SMM~1A and SMM~6.
\end{abstract}


\section{Introduction}\label{intro}
The earliest stages of stellar evolution for low-mass stars (0.2--2M$_\odot$) are becoming reasonably clear (see e.g. \citealp{2000prpl.conf...59A} for a review). The prestellar core stage \citep{1994MNRAS.268..276W} represents the phase in which a gravitationally bound molecular cloud core has formed. Thereafter gravitational collapse sets in (possibly mediated by turbulence or magnetic fields) and a central hydrostatic protostar forms, which together with its accreting disk and envelope, is known as a Class 0 protostar \citep{1993ApJ...406..122A}. Once half of the mass has accreted onto the central object it is known as a Class I protostar \citep{1989ApJ...340..823W}, and it subsequently evolves through the Class II \& III young stellar object (YSO) phases \citep{1987IAUS..115....1L}. To date only about 40 Class 0 protostars are known, and a similar number of prestellar cores \citep{init06}. Debate continues over the details of the evolutionary process, and submillimetre studies of more star-forming regions are required to help clarify the exact manner in which star formation proceeds.

The R Coronae Australis dark cloud is found at a distance of 170 pc \citep{1998A&A...338..897K} and is one of the closest star-forming regions to the sun. The cloud is centred on the Herbig Ae/Be stars RCrA and TCrA, which are the illumination sources of a bright reflection nebula (NGC 6729), and from which the region takes its name. There are also a number of other bright young stars in the region \citep[e.g.][]{1976AJ.....81..317V}. The cloud was previously believed to have an unusually low star-forming efficiency \citep{1976AJ.....81..317V,1981AJ.....86...62M}. The discovery of a cluster of young stars termed the Coronet \citep{1984MNRAS.209P...5T} close to RCrA, changed this view. The cloud is now known to be actively forming stars, and contains a number of embedded objects revealed through radio \citep{1987ApJ...322L..31B,1996AJ....111..320S} and millimetre wavelength observations \citep{2003A&A...409..235C,2004ApJ...600L..55C}.

There are a small number of nearby star-forming regions that we can study with good spatial resolution. If we are to understand the process of star formation, it is important that we characterise the similarities and differences between these regions. The RCrA cloud is particularly interesting, as it is not a member of the Gould belt. Its formation is therefore potentially dominated by different physics than in the majority of other nearby clouds, which have formed as part of the Gould belt. 

In this paper, we report on submillimetre observations of the RCrA molecular cloud. The observations are compared to those taken at other wavelengths, in an endeavour to understand the protostellar and prestellar population of the cloud.

\section{Observations}
\begin{table*}
\begin{center}
\caption{The positions of each of the maps made of the RCrA molecular cloud. The atmospheric conditions for each observation are indicated by the 225 GHz optical depth in column 7. The $1\sigma$ noise level measured in each region is given in columns 8 and 9.}
\label{obs_table}
\vspace{0.5cm}
\begin{tabular}{lcccccccc}\hline
	 &	    &			& \multicolumn{2}{c}{Map centre}&		&		   &\multicolumn{2}{c}{Mean $1\sigma$ noise level} \\ \cline{4-5}
Region 	 & Map      & Map Size  	& RA  		& Dec. 	   	& 	        &  		   &\multicolumn{2}{c}{(Jy/beam)} 	  \\ \cline{8-9}
Name	 & Type	    & (arcmin)		& (2000)	& (2000)	& UT Date	& $\tau_{225 GHz}$ & 850$~\mu$m & 450$~\mu$m \\ \hline
RCrA-A   & Scan     & 16$\times$11	& 19:01:50  	& $-$36:57:00	& 2000 Apr 09 	& 0.070     	& 0.041 & 2.6  \\ 
RCrA-B   & Scan     & 19$\times$9	& 19:02:50 	& $-$37:06:00	& 2000 Apr 10 	& 0.055     	& 0.030 & 1.7  \\ 
RCrA-C   & Scan     & 15$\times$10	& 19:03:50 	& $-$37:13:00	& 2000 Apr 11 	& 0.060     	& 0.031 & 2.2  \\ 
RCrA-A1A & Jiggle   & 2.3 		& 19:01:55	& $-$36:57:29	& 2000 Apr 11	& 0.065     	& 0.036	& 0.8	\\ 
RCrA-A1B & Jiggle   & 2.3 		& 19:01:48	& $-$36:55:10	& 2000 Apr 12	& 0.080     	& 0.033	& 1.2	\\ 
RCrA-A1C & Jiggle   & 2.3 		& 19:01:46	& $-$36:56:10	& 2000 Apr 12	& 0.080     	& 0.034	& 1.1	\\ 
RCrA-A1D & Jiggle   & 2.3 		& 19:01:42	& $-$36:58:10	& 2000 Apr 12	& 0.075     	& 0.032	& 1.0	\\ 
RCrA-A1E & Jiggle   & 2.3 		& 19:01:07	& $-$36:57:28	& 2000 Apr 12	& 0.075     	& 0.023	& 1.1	\\ 
RCrA-B1A & Jiggle   & 2.3 		& 19:02:58	& $-$37:07:35	& 2000 Apr 11	& 0.065     	& 0.018	& 0.8	\\ 
RCrA-B1B & Jiggle   & 2.3 		& 19:03:04	& $-$37:07:35	& 2000 Apr 12	& 0.075		& 0.022	& 0.8	\\ 
RCrA-B2A & Jiggle   & 2.3 		& 19:03:07	& $-$37:12:44	& 2000 Apr 11	& 0.065     	& 0.020	& 1.1	\\ 
RCrA-B2B & Jiggle   & 2.3 		& 19:03:00	& $-$37:14:05	& 2000 Apr 12	& 0.075     	& 0.022	& 0.8	\\ 
RCrA-B2C & Jiggle   & 2.3 		& 19:03:03	& $-$37:15:04	& 2000 Apr 12	& 0.075     	& 0.020	& 1.2	\\ 
RCrA-C1A & Jiggle   & 2.3 		& 19:04:00	& $-$37:15:31	& 2000 Apr 11	& 0.065     	& 0.015	& 0.4	\\ 
RCrA-C1B & Jiggle   & 2.3 		& 19:03:53	& $-$37:16:31	& 2000 Apr 12	& 0.075     	& 0.021	& 0.8	\\ \hline
\end{tabular}
\end{center}
\end{table*}

The observations were carried out using the Submillimetre Common User Bolometer Array (SCUBA) on the James Clerk Maxwell Telescope (JCMT). This instrument allows observations at 450 and 850$~\mu$m simultaneously through the use of a dichroic beam-splitter. The telescope has a resolution of 8 arcsec at 450$~\mu$m and 14 arcsec at 850 $\mu$m.

Observations of the RCrA cloud were carried out on the nights of 2000 April 9 -- 12 between 01:30 and 09:30 HST, using both scan-map and jiggle-map observing modes. In a jiggle-map, a 64-position jiggle pattern is used to Nyquist sample the 450 and 850$~\mu$m arrays, producing a map of diameter 2.3 arcmin.
A scan-map is made by scanning the array across the sky. The scan direction is 15.5$^\circ$ from the axis of the array in order to achieve Nyquist sampling. The array is rastered across the sky to build up a map several arcminutes in extent.

The cloud was scan-mapped in three sections, which we have labelled RCrA-A, B and C. In addition, jiggle-map observations were carried out of all of the sources that were detected in the scan-maps. The jiggle-maps have a larger integration time at each position and hence have better signal to noise. This allowed for significant detections of most sources at 450$~\mu$m. The observation details are summarised in Table~\ref{obs_table}. The scan-map areas are chosen to roughly trace the lowest contour of C$^{18}$O measured by \citet{1993A&A...278..569H}.

\begin{figure*}
\includegraphics[angle=0,width=160mm]{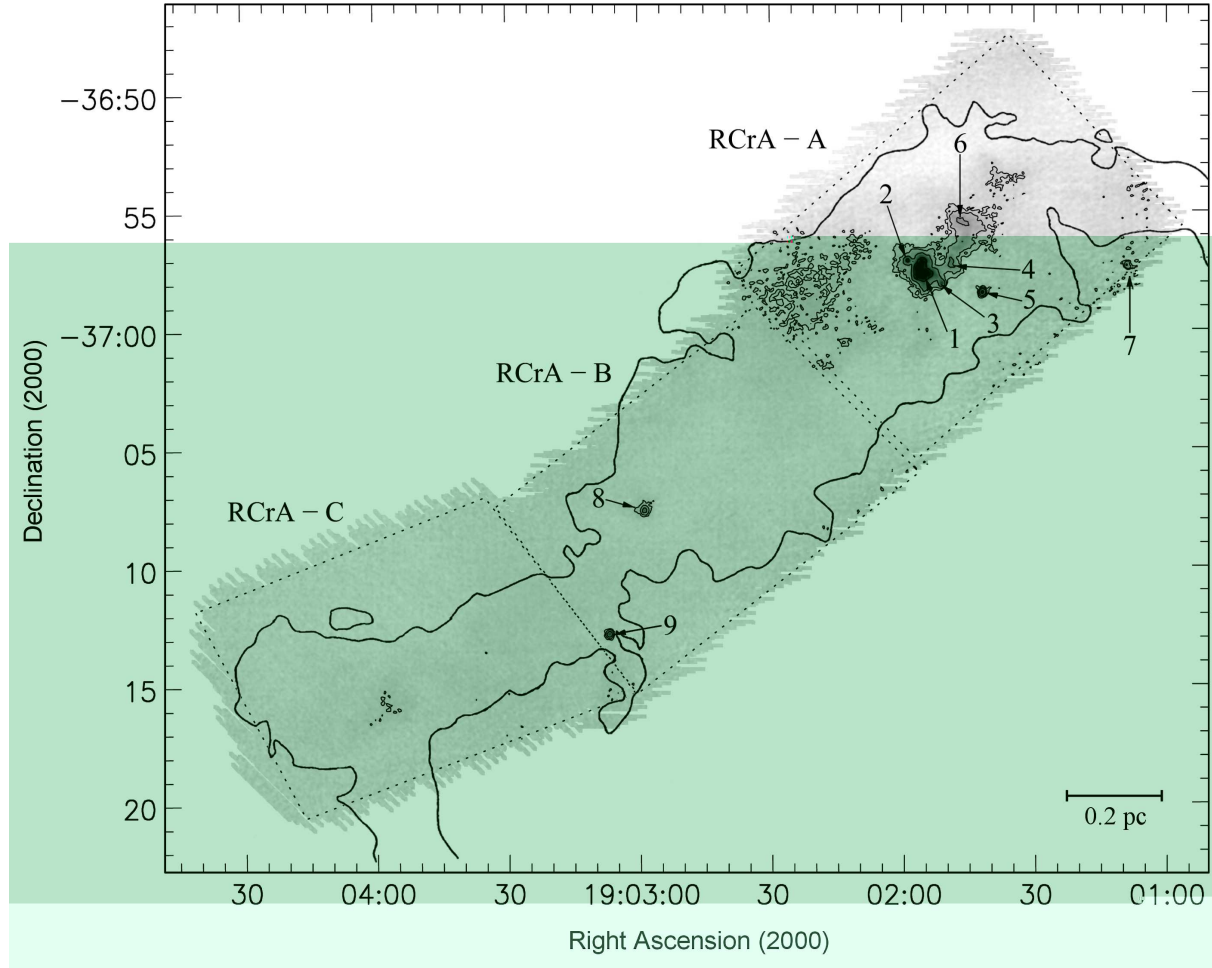}
\caption{The RCrA region seen at 850~$\mu$m. The contour levels are at 5$\sigma$, 10$\sigma$, 20$\sigma$, 40$\sigma$ and 60$\sigma$, where $\sigma$ is the mean noise level over the whole region, equal to 0.035 Jy/beam. The boundaries of the scan-maps are shown as dotted lines. The identified cores SMM~1--9 are marked on the map. The outermost bold contour delineates the 1~Kkms$^{-1}$ contour from the C$^{18}$O ($J=1-0$) of \citet{1993A&A...278..569H}. }
\label{scan-map}
\end{figure*}

\begin{table*}
\begin{center}
\caption{The position and measured flux density of each core in the RCrA maps at 450 and 850~$\mu$m. The name(s) of the jiggle-map(s) covering each source is given in column 4. Column 5 indicates whether or not the core is extended in the 8-arcsec 450-$\mu$m JCMT beam. The peak flux density in Jy/beam is quoted in columns 7 and 9. In addition the integrated flux density within an aperture is given for extended sources (columns 8 \& 10). The aperture size used is given in column 6. SMM~1B \& 1C are unresolved in the raw 850~$\mu$m data. *The flux density from the extended sources SMM~4 \& 6 are given in columns 8 \& 10 in units of Jy/24-arcsec beam for comparison with 1.2mm data at that resolution (see text for details).}
\label{fluxes}
\vspace{0.5cm}
\begin{tabular}{lcclcccccc}\hline
 	&  		&  		&			&  		& Aperture 		& \multicolumn{2}{c}{850$~\mu$m} & \multicolumn{2}{c}{450$~\mu$m} \\ \cline{7-10}
Source 	& RA 		& Dec. 		& \multicolumn{1}{c}{Jiggle-map}&Extended& Diam	 		& Peak		& Int. 		& Peak	  	& Int.		\\
Name	&(2000)		&(2000)		& \multicolumn{1}{c}{Name}&   		& (arcsec)		& (Jy/beam)	& (Jy)		&(Jy/beam)	&(Jy)		\\ \hline
SMM~1A 	& 19:01:55.6 	& $-$36:57:43 	& ~~~A1A		& Y 	  	& $47\times 25$		& 2.9	  	& 8.1    	& 26		& 120		\\
SMM~1B 	& 19:01:56.3 	& $-$36:57:30 	& ~~~A1A		& Y 	  	& $18\times 17$		& 3.1	  	& 5.4    	& 22		& ~50		\\
SMM~1C 	& 19:01:55.4 	& $-$36:57:20	& ~~~A1A		& Y	  	& $16\times 16$		& $<2.6$	& 5.6	   	& 16		& ~45 	\\
SMM~2 	& 19:01:58.8 	& $-$36:57:07	& ~~~A1A		& N 	  	& --		       	& 1.5		& --    	& 10	 	& --		\\
SMM~3 	& 19:01:51.1 	& $-$36:58:04 	& ~~~A1A		& N 	  	& --	 		& 1.5		& --    	& ~9	 	& --		\\
SMM~4	& 19:01:49.0 	& $-$36:57:11 	& ~~~A1A/A1C		& Y 	  	& $24\times24$*		& 0.8	  	& 1.8		& --		& ~12		\\
SMM~5	& 19:01:41.9 	& $-$36:58:27 	& ~~~A1D		& N 	  	& --	 		& 2.0		& --    	& 12	 	& --		\\
SMM~6 	& 19:01:46.4 	& $-$36:55:30 	& ~~~A1B/A1C		& Y	  	& $24\times24$*		& 0.9	  	& 2.0		& --		& ~~9		\\
SMM~7 	& 19:01:09.1 	& $-$36:57:19 	& ~~~A1E		& N 	  	& --	 		& 0.7		& --	   	& ~3	 	& --		\\
SMM~8	& 19:02:58.9 	& $-$37:07:37 	& ~~~B1A/B1B		& N 	  	& --	 		& 1.1		& --    	& ~9	  	& --		\\
SMM~9	& 19:03:06.9 	& $-$37:12:50 	& ~~~B2A		& N 	  	& --	 		& 1.5		& --    	& 12		& --		\\ \hline
\end{tabular}
\end{center}
\end{table*}

Time-dependent variations in the sky emission were removed by chopping the secondary mirror at 7.8 Hz. For the jiggle-maps, a chop throw of 150 arcsec in azimuth was used. The scan-maps are too large for the chop position to be outside of the map, and still successfully remove sky variations. A chop throw much smaller than the map size is used. Each source in the map therefore appears as a positive and a negative source. The region is mapped six times, with chop throws of 30, 44 and 68 arcsec in both RA and Dec \citep{1995mfsr.conf..309E}. The dual-beam function is removed from each map in Fourier space by dividing each map by the Fourier transform of the dual-beam function, which is a sinusoid. The multiple chop-throws allow for cleaner removal of the dual beam function in Fourier space. The maps are them combined, weighting each map to minimise the noise introduced at the spatial frequencies that correspond to zeroes in the sinusoids. Finally the map is converted back into normal space, and no longer contains the negative sources \citep{SURF}.

The submillimetre zenith opacity at 450 and 850$~\mu$m was determined using the `skydip' method and by comparison with polynomial fits to the 1.3~mm sky opacity ($\tau_{225 GHz}$) \citep{2002MNRAS.336....1A} measured at the Caltech Submillimeter Observatory. The average 850$~\mu$m zenith optical depth was 0.27, corresponding to a mean zenith transmission of $\sim 75$\%. The average 450$~\mu$m optical depth was 1.44, corresponding to a zenith transmission of $\sim 25$\%. The telescope pointing was checked at regular intervals throughout the nights using planets, secondary calibrators and standard pointing sources. 

The data were reduced in the normal way using the SCUBA User Reduction Facility \citep{SURF}. Calibration was performed using observations of the planet Uranus taken during each shift. We estimate that the absolute calibration uncertainty is $\pm$10\% at 850$~\mu$m and $\pm$30\% at 450$\mu$m, based on the consistency and reproducibility of the calibration. 

\section{Results}

\begin{figure*}
\includegraphics[angle=0,width=0.79\textwidth]{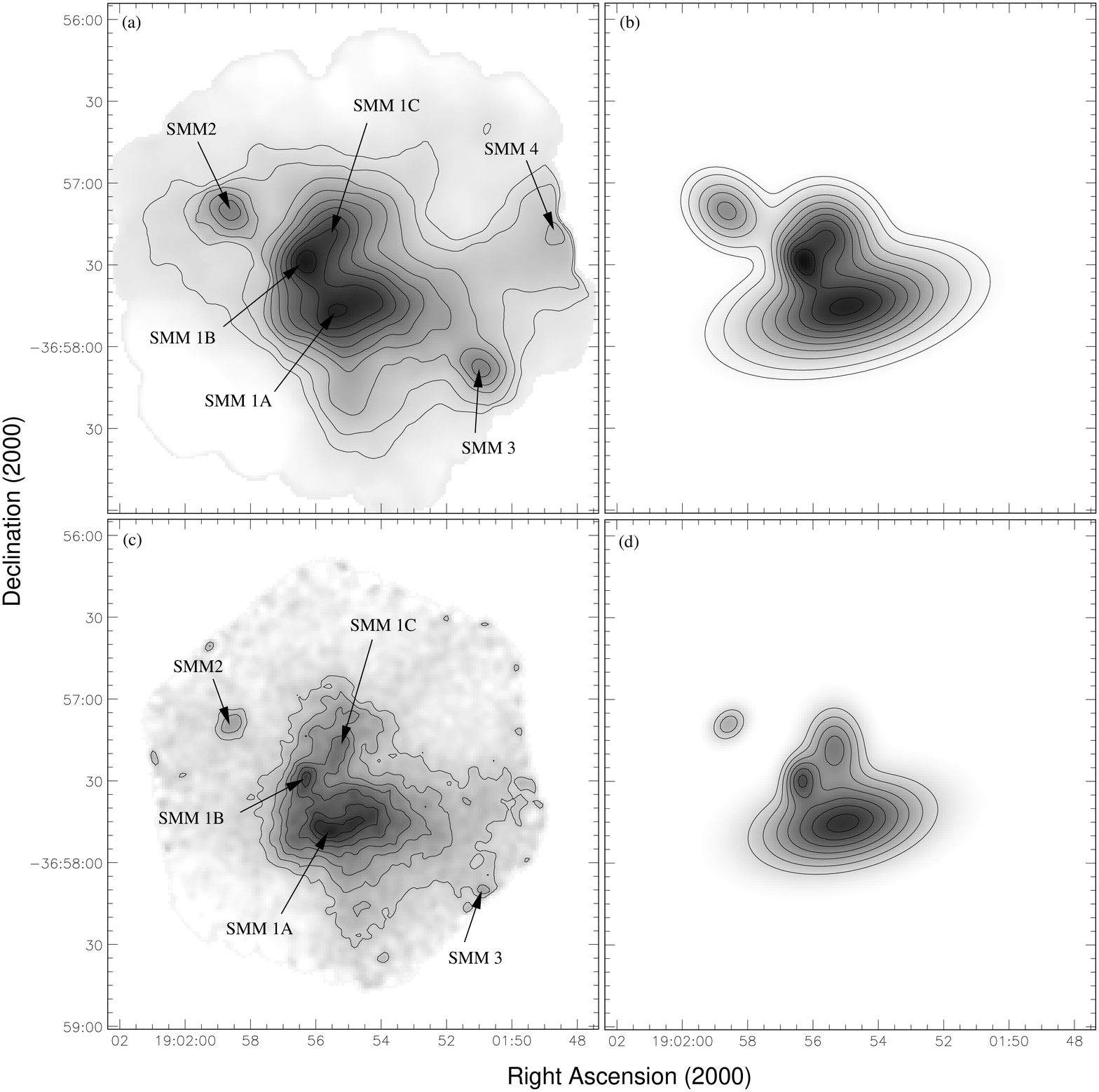}
\caption{The RCrA-A1A jiggle-map, showing a close-up view of SMM~1 at (a) 850~$\mu$m and (c) 450~$\mu$m. The components of SMM~1 and other nearby sources are indicated. The results of the Gaussian composite fitting of SMM~1 \& 2 to the 850 and 450~$\mu$m maps are shown in (b) and (d) respectively.}
\label{models}
\end{figure*}

Figure~\ref{scan-map} shows the scan-map data for the RCrA region at 850$~\mu$m.  The area covered is approximately 0.15 deg$^2$, which at a distance of 170 pc is equal to $\sim$ 1.4 pc$^2$. The RCrA-A, B and C scan-map boundaries are marked on Figure~\ref{scan-map}. The positions of the jiggle-maps are listed in Table 1. The 1$\sigma$ noise level in each of these maps is also given in Table~\ref{obs_table}. 

A number of features can be seen in the region, some of which are reasonably low level and extended whilst others are more point-like. We detect nine roughly point-like sources (although one source is subsequently seen to contain three components -- see below). We number the sources in order of increasing distance from the brightest part of RCrA. The positions of the sources are listed in Table~\ref{fluxes}.

We have labelled the brightest feature in the submillimetre SMM~1. The jiggle-maps (RCrA-A1A) of this source at 450 and 850~$\mu$m are shown in Figure~\ref{models}. When examined closely, SMM~1 can be seen to break up into three distinct components, which we have labelled SMM~1A, 1B and 1C.

SMM~1 is surrounded by a number of other sources (SMM~2,3 \& 5). A filament extends to the north-west, containing two extended flux density peaks, which we label SMM~4 \& 6. SMM~7 lies to the west of the main clump. To the east of the main clump is an area of increased flux density, though no point sources are detected within it.

The RCrA-B and RCrA-C regions show much less structure. RCrA-B contains two unresolved sources, which we label SMM~8 \& 9. The only structure in RCrA-C is an area of increased flux density containing no significant point sources. All sources are detected at 450$~\mu$m, with the exception of SMM~4 \& 6.  

Table~\ref{fluxes} gives the measured flux density for each object. The peak flux density is given in Jy/beam. SMM~4 and 6 are not detected at 450$~\mu$m without smoothing the data. For this reason, and for comparison with SIMBA data at a resolution of 24 arcsec (see Section~\ref{discussion_other}), the flux density from these objects is quoted in columns~8 \& 10 of Table~\ref{fluxes} in units of Jy/24-arcsec beam at both 450 and 850$~\mu$m. At this resolution, the noise levels at 850~$\mu$m and 450~$\mu$m are 0.030~Jy/beam and 0.54~Jy/beam respectively.

SMM~1A, 1B and 1C have a degree of overlap on the plane of the sky, therefore there will be some flux density contamination in each aperture from the other two objects. To account for this, the flux densities of SMM~1A, 1B, 1C and the adjacent source SMM~2 were modelled as a collection of Gaussians. The Gaufit routine \citep{ESP} was used, which fits a number of 2-D Gaussians to an image. The dimensions and positions of the Gaussians were left as free parameters. The 450 and 850~$\mu$m data were fitted independently. The results of the fitting are shown in Figure~\ref{models}(b) and (d), and listed in Table~\ref{model_table}. The parameters of each fit were used to calculate the integrated flux density for each source, and these are also listed in Table~\ref{model_table}. 

When the integrated flux density calculated in this way is compared to the directly measured value (see Table~\ref{fluxes}), it is clear that the measured flux densities for SMM~1B and 1C are increased as a result of the nearby extended source SMM~1A. SMM~1B receives a larger contribution as a result of being closer to SMM~1A on the plane of the sky. The deconvolved FWHMs at 450 and 850 $\mu$m are consistent for each source.

\begin{table}
\begin{center}
\caption{The parameters of the Gaussian fitting for the three components of SMM~1 and for SMM~2.}
\label{model_table}
\vspace{0.5cm}
\begin{tabular}{lccc}\hline
Source	& FWHMa		& FWHMb		&Integrated flux\\
Name	& (arcsec)	& (arcsec)	&density (Jy)	\\ \hline
	 \multicolumn{4}{c}{450$~\mu$m} 		\\ \hline 
SMM~1A	& 47.4		& 23.6		& 157.0		\\
SMM~1B	& 13.6		& 10.3		& 13.6		\\
SMM~1C	& 22.4		& 16.6		& 33.2		\\ 
SMM~2	& 12.2		&  9.5		& 6.8		\\ \hline
	 \multicolumn{4}{c}{850$~\mu$m} 		\\ \hline
SMM~1A	& 53.2		& 24.0		& 14.4		\\
SMM~1B	& 16.1		& 11.4		& 1.4 		\\
SMM~1C	& 22.2		& 23.8		& 4.0 		\\ 
SMM~2	& 20.8		& 16.2		& 1.9		\\ \hline
\end{tabular}
\end{center}
\end{table}

\section{Discussion}\label{discussion}
\subsection{SMM~1}\label{discussion_smm1}
\begin{figure}
\includegraphics*[angle=0,width=83mm]{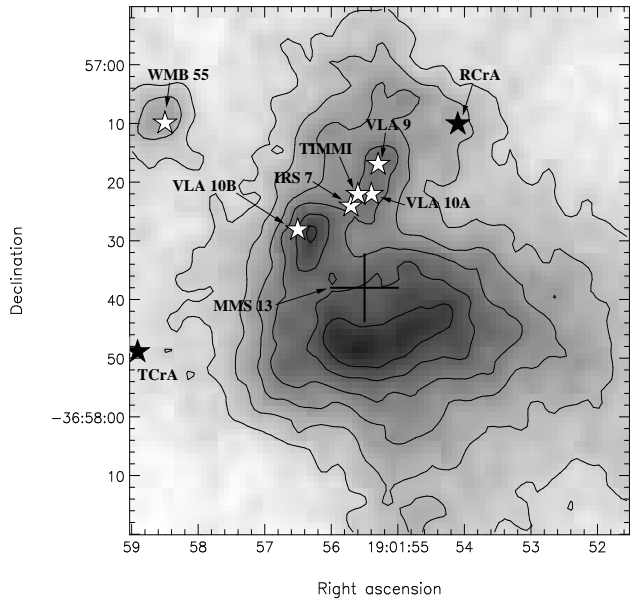}
\caption{The immediate vicinity of SMM~1 at 450~$\mu$m. The contour levels are at 5$\sigma$, 10$\sigma$, 15$\sigma$, 20$\sigma$, 25$\sigma$ and 30$\sigma$, where $\sigma$ is equal to 0.76 Jy/beam. The positions RCrA and TCrA are marked with filled stars. In addition, the positions of other sources from the literature are marked with open stars as follows:  IRS~7 \citep{1984MNRAS.209P...5T}, VLA~9, VLA~10A, VLA~10B \citep{1987ApJ...322L..31B}, WMB~55 and TIMMI \citep{1997AJ....114.2029W}. MMS~13 \citep{2003A&A...409..235C} is marked with a cross equal in size to the HWHM of the 1.2~mm data of \citeauthor{2003A&A...409..235C}. VLA~10A and TIMMI are claimed to be the same source \citep{1997AJ....114.2029W}.}
\label{composite}
\end{figure}

The source SMM~1 is the brightest source in the map, and as discussed above, appears to be composed of three objects. Observations at different wavelengths of the vicinity of this source have revealed a number of different objects, the nature of which has been the subject of much debate. Figure~\ref{composite} shows the 450~$\mu$m jiggle-map data of SMM~1. The positions of other sources from the literature are marked.

In the environs of SMM~1 lies the IRS~7 source, which was first detected in the near infrared at $\lambda = 2 \mu m$ \citep{1984MNRAS.209P...5T}. It was characterised as a very young star embedded in a warm dust shell. VLA observations of the region revealed two centimetre sources close to IRS~7 \citep{1987ApJ...322L..31B}. The two cm sources are approximately equi-distant from the IRS~7 infrared source, and were interpreted as arising from the interaction between a stellar wind from IRS~7 and a thick accretion disk. The two cm sources (VLA~10A and VLA~10B) are labelled on Figure~\ref{composite}. Also shown is the position of the nearby source VLA~9 \citep{1987ApJ...322L..31B}.

\citet{1997AJ....114.2029W} observed the region at 10$~\mu$m using the TIMMI camera at the ESO 3.6 m telescope. They discovered a 10$~\mu$m source which is labelled in Figure~\ref{composite} as TIMMI. This source lies much closer to VLA~10A than to VLA~10B. \citeauthor{1997AJ....114.2029W} claim that the near infrared detection at the original position of IRS~7 is most likely to be reflection nebulosity, and they therefore refute the conclusion about the relationship between IRS~7, VLA~10A and VLA~10B made by \citet{1987ApJ...322L..31B}. Instead, they claim that VLA~10A and the TIMMI are the same deeply embedded protostar, while they state that VLA~10B is of an unknown nature. Our source SMM~1C is coincident with the VLA~10A/TIMMI object. We therefore concur with \citet{1997AJ....114.2029W} that VLA~10A/TIMMI/SMM~1C is a deeply embedded protostar. We also note that our source SMM~1C is extended in the direction of VLA~9. It is therefore possibly a composite of these two objects. 

SMM~1B is coincident with VLA~10B. We conclude that SMM~1B is most likely to be a Class 0 protostar \citep{1993ApJ...406..122A}. This conclusion is based on the object's large submillimetre flux density, which traces a concentration of gas and dust, and the cm emission from VLA~10B, which reveals the presence of a protostellar object, embedded at the centre of the core. In addition, there have been no near or mid-infrared sources detected, which would indicate the presence of a more evolved protostar. This conclusion could be confirmed by the detection of a collimated outflow, centred on SMM~1B. 

This is the first resolved detection of the dust continuum radiation from this protostar. Previous studies have led to the identification a Class 0 protostar in this region \citep{1996A&A...309..827S,2003A&A...409..235C}, but have lacked the spatial resolution to disentangle the emission from the protostar from that of nearby sources.

We have compared our 450 and 850~$\mu$m data to 200~$\mu$m data taken with the ISOPHOT camera \citep{1996A&A...315L..64L} on-board the Infrared Space Observatory (ISO) \citep{1996A&A...315L..27K}. The observations were made using the over-sampled mapping mode PHT32 -- cf. \citet{2002MNRAS.329..257W}. For more details, see \citet{PHT_handbook}. The data were obtained through the ISO data archive \citep{1996A&A...315L..27K}. 

\begin{figure}
\includegraphics*[angle=0,width=83mm]{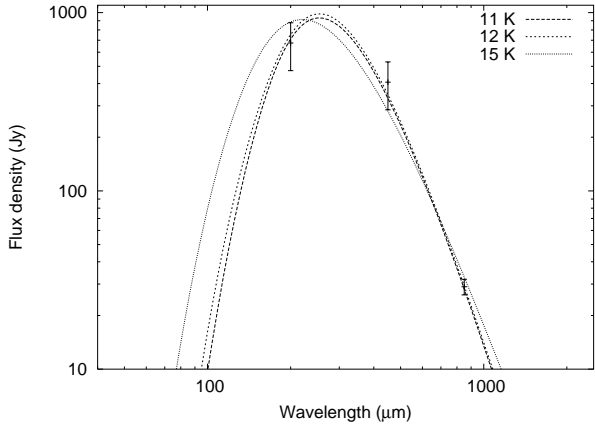}
\caption{The spectral energy distribution for SMM~1, 2 and 3. Error bars of 30\% at 200~$\mu$m, 30\% at 450~$\mu$m and 10\% at 850~$\mu$m are plotted. The best fitting greybody curves for 11, 12 and 15~K are also plotted, see text for discussion.}
\label{SMM1_SED}
\end{figure}

At a wavelength of 200~$\mu$m, ISO has an angular resolution of 84 arcsec. The sources SMM~1, 2 and 3 are therefore unresolved in the ISO beam. In order to compare the SCUBA data with the 200 $\mu$m ISOPHOT data, the SCUBA data were smoothed to the same resolution as the ISOPHOT data. The integrated flux density was then measured for each map in an aperture of diameter 160 arcsec. This aperture encompasses SMM~1, 2 and 3. The flux densities at 200, 450 and 850~$\mu$m listed in Table \ref{smoothed_flux}. The spectral energy distribution (SED) based on these data is plotted in Figure \ref{SMM1_SED}. The data are plotted with error bars of 30\% at 200~$\mu$m, 30\% at 450~$\mu$m and 10\% at 850~$\mu$m.  

Figure \ref{SMM1_SED} also shows a number of modified blackbody or greybody curves of the form:
\begin{equation}
F_{\nu} = B_{\nu,T}(1-e^{-\tau_\nu})\Omega
\label{eqn_greybody}
\end{equation}
where $F_{\nu}$ is the flux density at frequency $\nu$, $B_{\nu,T}$ is the Planck function at frequency $\nu$ and temperature T, $\Omega$ is the solid angle subtended by the source and $\tau_\nu$ is the optical depth at frequency $\nu$ which is given by $\tau_\nu \propto \nu^\beta$.

\begin{table}
\begin{center}
\caption{The flux density of SMM~1, 2 and 3 at 200, 450 and 850~$\mu$m, measured in an aperture of diameter 160 arcsec}
\label{smoothed_flux}
\vspace{0.5cm}
\begin{tabular}{lcc}\hline
Instrument 	& Wavelength 	& Flux Density	\\
		& ($\mu$m)	& (Jy)		\\ \hline
ISOPHOT		& 200		& 675		\\
SCUBA		& 450		& 407		\\
SCUBA		& 850		& 29		\\ \hline
\end{tabular}
\end{center}
\end{table}

The parameter space of Equation \ref{eqn_greybody} was explored to determine the range of temperatures that will give acceptable fits to the measured flux densities at the three wavelengths. The small number of points in the SED mean that the parameters of the greybody cannot be uniquely determined. 

The temperature can be constrained be restricting the solid angle $\Omega$. A lower limit for the radius of $\Omega$ was set by the measured source HWHM, which after deconvolving the telescope beam, is equal to 20 arcsec. This sets an upper limit of 15~K for the temperature of a greybody that gives an acceptable fit to the data. If a conservative upper limit for the radius of $\Omega$ of twice the measured source HWHM is imposed, then a lower limit of 11~K is set for the temperature. The best fitting grey-body within this range has a temperature of 12~K. 

Greybody fits for the two extremes of this temperature range are shown in Figure~\ref{SMM1_SED}, along with the best fitted greybody. The parameters of these fits are given in Table \ref{greybody_parameters}. We use this temperature range when determining the envelope mass from the 850~$\mu$m flux density. This is given in Table~\ref{masses}.

\begin{table}
\begin{center}
\caption{The parameters of the three greybody fits shown in Figure \ref{SMM1_SED}. Column 2 gives the radius of the solid angle. Column 4 gives the critical wavelength where the optical depth is equal to~1.}
\label{greybody_parameters}
\vspace{0.5cm}
\begin{tabular}{cccc}\hline
Temp 	& $\Omega$ 	& $\beta$	& $\lambda_c$	\\
(K)	& (arcsec)	& 		& ($\mu$m)	\\ \hline
11	& 40		& 3.4		& 240		\\
12	& 30		& 3.3		& 265 		\\
15	& 20		& 2.5		& 223		\\ \hline
\end{tabular}
\end{center}
\end{table}

The SED fitting is based on the combined data of SMM~1, 2 and 3. However at both 450 and 850 $\mu$m, $>70$\% of the flux density originates from SMM~1A. We therefore believe that the temperature obtained in this manner is indicative of the temperature of this source. As discussed above, we believe sources SMM~1B and 1C to be protostellar in nature and therefore assume a canonical temperature of 30~K for these sources \citep[e.g.][]{1993ApJ...406..122A}.

\begin{table*}
\begin{center}
\caption{The prestellar core and protostellar envelope masses for the SCUBA detections in the RCrA map (see text for details). References in column~5 are:$^{1}$\citealp{2003A&A...409..235C}; $^{2}$\citealp{1987ApJ...322L..31B}; $^{3}$\citealp{1997AJ....114.2029W}; 
$^{4}$\citealp{1984MNRAS.209P...5T}; $^{5}$\citealp{1974ApJ...187...83S}; $^{6}$\citealp{1945ApJ...102..168J}; 
$^{7}$\citealp{1999A&A...350..883O}; $^{8}$\citealp{1992ApJ...397..520W}; $^{9}$\citealp{1972ApJ...174..401H}.}
\label{masses}
\vspace{0.5 cm}
\begin{tabular}{lcccl}\hline
Source 	& Assumed	  & Class 	& Envelope Mass	& Associations 			\\ 
Name 	& Temperature (K) & 		& (M$_\odot$)	& 				\\ \hline
SMM~1A	& $11 - 15$	  & ~Pre	& $6.4 - 11.2$	& MMS~13$^{1}$			\\
SMM~1B	& 30		  & ~0		& 0.2 		& VLA~10B$^{2}$			\\
SMM~1C	& 30		  & ~I		& 0.6		& VLA~10A$^{2}$, TIMMI$^{3}$	\\
SMM~2	& 30		  & ~I		& 0.3		& WMB~55$^{1,3}$		\\
SMM~3	& 30		  & ~I		& 0.3		& IRS~1$^{4}$, HH100 IRS$^{5}$, VLA~3$^{2}$	\\
SMM~4	& 30		  & ~I 		& 0.3		& IRS~5$^{4}$, MMS~10$^{1}$, VLA~7$^{2}$	\\
SMM~5	& 30		  & ~I		& 0.3		& IRS~2$^{4}$, MMS~9$^{1}$	\\
SMM~6	& $10 - 15$	  & ~Pre	& $0.9 - 1.9$	& MMS~10$^{1}$ 			\\
SMM~7	& 30		  & ~II		& 0.1		& SCrA$^{6}$, MMS~1$^{1}$, ISO-CrA-116$^7$  \\
SMM~8	& 30		  & ~I		& 0.2 		& IRAS~18595-3712$^{8}$, MMS~23$^{1}$, IRAS~32c$^{8}$ \\ 
SMM~9	& 30		  & ~I 		& 0.2		& VVCrA$^{9}$, MMS~24$^{1}$			\\ \hline
\end{tabular}
\end{center}
\end{table*}

In Table~\ref{masses} we summarise the results and give mass estimates of the prestellar cores and protostellar envelopes for all of our detected sources. The envelope mass $M$ is calculated using the following:
\begin{equation}
M=\frac{S_{850}D^2}{\kappa_{850}B_{850,T}},
\end{equation}
where $S_{850}$ is the 850~$\mu$m flux density, $D$ is the distance to the source, $\kappa_{850}$ is the mass opacity of the gas and dust, and $B_{850,T}$ is the Planck function at temperature $T$. We follow \citet{1996A&A...314..625A} who use a 1.3~mm mass opacity of ${\rm 0.005~cm^2g^{-1}}$ which assumes a gas to dust ratio of 100 \citep[e.g.][]{1983QJRAS..24..267H}. We scale this value to 850~$\mu$m using the canonical value of $\beta$ of 1.5, which yields a value of $\kappa_{850}$ of ${\rm 0.01~cm^2g^{-1}}$. We do this to be consistent with previous work \citep[e.g.][]{1996A&A...314..625A,1999MNRAS.305..143W,init06}. However, we note that for SMM~1, we have measured a $\beta$ of $2.5-3.5$. This would decrease the measured masses for the components of SMM~1 quoted in Table \ref{masses} by $\sim 2$. We note that we have been unable to accurately determine $\beta$ for the other sources.

The 1.2 mm continuum map of \citet{2003A&A...409..235C} shows a similar morphology to our SCUBA data, featuring a strong peak --- named MMS~13 by \citeauthor{2003A&A...409..235C} --- between RCrA and TCrA. The position of MMS~13 is marked on Figure~\ref{composite}. The position and morphology of MMS~13 is very similar to that of SMM~1 in both the 450 and 850 $\mu$m SCUBA data, when they are smoothed to 24 arcsec resolution. \citeauthor{2003A&A...409..235C} conclude that MMS~13 may be a deeply embedded Class~0 source, or possibly a proto-Herbig-Ae/Be star.

Our higher resolution data show that MMS~13 is a composite of SMM~1A, 1B and 1C. As discussed above, we agree that SMM~1B and 1C are probably protostellar. However, we believe that SMM~1A may be prestellar in nature, due to the lack of a detection of a centimetre source coincident with SMM~1A, and the low temperature obtained from the SED fitting. The low temperature indicates the lack of a central heating source.

To summarise, we believe that there are at least three sources in the immediate vicinity of IRS~7. The cm source VLA~10A is the least embedded, and is the only source visible in the infrared. The extended envelope that is detected in the submillimetre, together with the presence of a mid-infrared source, leads us to believe that SMM~1A is probably a Class~I object. VLA~10B has only been detected and resolved from nearby sources at cm wavelengths \citep{1987ApJ...322L..31B} and in our SCUBA data (SMM~1B). This object is probably a Class~0 protostar, as discussed above. The third source (SMM~1A) is only detected in the submillimetre and at 1.2 mm \citep{2003A&A...409..235C}, and we believe it to be prestellar in nature.

\subsection{Other Sources}\label{discussion_other}
The majority of the other submillimetre sources in the region correspond to known young stellar objects (YSOs). SMM~2 was detected at 2$~\mu$m by \citet{1997AJ....114.2029W} though it was not identified as an association member. It is also seen as a shoulder on the side of MMS~13 in the 1.2 mm map of \citet{2003A&A...409..235C}, who suggested that it is a Class~I object. SMM~2 is shown in Figure~\ref{models}(a) and (c), and also in Figure~\ref{composite} labelled as `WMB~55'. We have too few data-points to constrain the SED for this source, therefore cannot determine the dust temperature. We concur with \citet{2003A&A...409..235C} that the source is probably a Class~I protostar and therefore use a canonical dust temperature of 30~K to determine the envelope mass (see Table~\ref{masses}).

SMM~3 corresponds to the Coronet member IRS~1 \citep{1984MNRAS.209P...5T}, which is also known as HH100~IRS. This is the driving source of the Herbig-Haro object HH100 \citep{1974ApJ...187...83S}, which was first discovered by \citet{1974ApJ...191..111S}. It has also been detected at a wavelength of 6~cm (source VLA~3 of \citealp{1987ApJ...322L..31B}). This source is seen as a shoulder on the side of MMS~13, though it is not resolved in the 1.2 mm data \citep{2003A&A...409..235C}. HH100~IRS has a rising SED between 2 and 25~$\mu$m \citep{1984MNRAS.209P...5T,1992ApJ...397..520W}, indicating that it is a Class~I protostar \citep{1987IAUS..115....1L}. We therefore assume a dust temperature of 30~K when calculating the envelope mass. This is quoted in Table~\ref{masses}.

The source SMM~4 is found in the filament stretching from SMM~1 to the north-west, and is notably extended in the 14-arcsec JCMT beam. SMM~4 coincides with the Coronet member IRS~5 \citep{1984MNRAS.209P...5T}. It has also been detected at 1.2~mm (labelled MMS~12 by \citealp{2003A&A...409..235C}), and at 6~cm (labelled VLA~7 by \citealp{1987ApJ...322L..31B}). SMM~4 was not detected in the IRAS survey of the region \citep{1992ApJ...397..520W}. Consequently the mid-infrared slope of its SED is not known. Thus, the submillimetre and millimetre data do not constrain the dust temperature. Its evolutionary status is therefore unclear. Based on an interpolation between the near-infrared and submillimetre flux densities, we assign this source as a Class I protostar, and therefore assume a temperature of 30~K.

SMM~5 coincides with the Coronet member IRS~2 \citep{1984MNRAS.209P...5T,1997AJ....114.2029W}. This source has been detected at 1.2~mm (labelled MMS~9 by \citealp{2003A&A...409..235C}). Mid-infrared observations show that it has a rising SED to longer wavelengths \citep{1992ApJ...397..520W}, indicating that this is a Class~I protostar. 

SMM~6 is the least centrally condensed source in our maps. The only other detection of this source is at 1.2 mm (labelled MMS~10 by \citealp{2003A&A...409..235C}).  \citeauthor{2003A&A...409..235C} claim that MMS~10 is probably a deeply embedded protostar. We argue that there is no indication of an embedded protostar (e.g. infrared or cm source), and therefore hypothesise that the object is prestellar in nature. The millimetre and sub-millimetre data do not constrain the temperature of this object. Therefore we calculate the dust mass for a temperature range of 10--15~K (typical for prestellar cores -- \citealp{2002MNRAS.329..257W}).

SMM~7 corresponds to the T-Tauri star S~CrA \citep{1945ApJ...102..168J}. It is a strong source at millimetre wavelengths \citep{1993A&A...273..221R, 2003A&A...409..235C}, and is the driving source of the Herbig-Haro object HH 82 \citep{1986ApJS...62...39S}. It is known to be a visual binary with a separation of 1.4 arcsec. It has been detected in the near and mid-infrared \citep{1992ApJ...397..520W,1999A&A...350..883O}. The slope of its SED between 2 and 25~$\mu$m indicates that it is a Class II YSO. The envelope dust temperature for Class II sources is generally taken to be 30~K \citep{1994ApJ...420..837A}. This temperature is consistent with greybody fits (Equation~\ref{eqn_greybody}) to the far-infrared \citep{1992ApJ...397..520W}, submillimetre (this study) and millimetre \citep{2003A&A...409..235C} flux densities. These data constrain the temperature to between 30 and 40~K. Its mass is quoted in Table~\ref{masses} under the assumption of a temperature of 30~K.

SMM~8 coincides with the source IRAS 18595-3712, which has also been detected in the near and mid-infrared \citep{1992ApJ...397..520W,1999A&A...350..883O}, and at millimetre wavelengths \citep{2003A&A...409..235C}. It has the mid-infrared SED of a Class I protostar. We therefore assume a dust temperature of 30~K \citep{1994ApJ...420..837A}. As for SMM~7, this temperature is consistent with greybody fits to the far-infrared, submillimetre and millimetre data.

SMM~9 corresponds to the double star VV~CrA \citep{1972ApJ...174..401H}, both components of which display T-Tauri behaviour \citep{1997ApJ...474..455P}. The slope of its mid-infrared SED is approximately flat, possibly indicating a YSO with an active or flared disk \citep{1988ApJ...326..865A}. The measured flux density from this source in the far-infrared \citep{1992ApJ...397..520W}, submillimetre (this study) and millimetre \citep{2003A&A...409..235C} can be fitted with greybody curves corresponding to a temperature range of $25 - 40$~K. We therefore use the canonical 30~K when calculating the envelope mass for this source.

\section{Conclusions}
We have mapped the R Coronae Australis molecular cloud at 450 and 850~$\mu$m, and have
compared these data to previous surveys of the region made at infrared, millimetre and centimetre wavelengths. We have discovered a new Class~0 object. This object is associated with a cm source which was previously of an unknown nature. The protostar is separated by 0.01 pc from a second source that we hypothesise to be prestellar in nature. In addition, we have determined the envelope masses of a number of previously identified YSOs and identified one further prestellar core.

\section*{Acknowledgments}
The authors would like to thank the staff of the JCMT for assistance with the observations. The JCMT is operated by the Joint Astronomy Centre, Hawaii, on behalf of the UK PPARC, the Netherlands NWO, and the Canadian NRC. DJN acknowledges PPARC for PDRA support.

\end{document}